\documentclass[a4paper,UKenglish,cleveref, autoref, thm-restate]{lipics-v2021}

\hideLIPIcs  

\title{Online Knapsack Problems with Estimates} 

\author{Jakub Balab\' an}{Masaryk University, Czechia}{jakbal@mail.muni.cz}{https://orcid.org/0000-0002-2475-8938}{Brno Ph.D. Talent Scholarship Holder – Funded by the Brno City Municipality}
\author{Matthias Gehnen}{RWTH Aachen University, Germany}{gehnen@cs.rwth-aachen.de}{https://orcid.org/0000-0001-9595-2992}{Funded by RWTH Research ENHANCE.R}
\author{Henri Lotze}{RWTH Aachen University, Germany}{lotze@cs.rwth-aachen.de}{https://orcid.org/0000-0001-5013-8831}{}
\author{Finn Seesemann}{RWTH Aachen University, Germany}{finn.seesemann@rwth-aachen.de}{https://orcid.org/0009-0002-9614-7738}{}
\author{Moritz Stocker}{ETH Zürich, Switzerland}{moritz.stocker@inf.ethz.ch}{}{}

\authorrunning{J. Balabán, M. Gehnen, H. Lotze, F. Seesemann, M. Stocker} 

\Copyright{Jakub Balabán, Matthias Gehnen, Henri Lotze, Finn Seesemann, Moritz Stocker} 



\category{} 

\relatedversion{} 

\nolinenumbers

\EventEditors{John Q. Open and Joan R. Access}
\EventNoEds{2}
\EventLongTitle{arxiv}
\EventShortTitle{arxiv 2025}
\EventAcronym{arxiv}
\EventYear{2025}
\EventDate{ 2025}
\EventLocation{Aachen, Germany}
\EventLogo{}
\SeriesVolume{}
\ArticleNo{}

\usepackage{pgfplots}
\usepgfplotslibrary{fillbetween}
\pgfsetlayers{background,edgelayer,nodelayer,main}
\usepackage{tikz}
\usetikzlibrary{matrix,arrows,backgrounds} 
\usetikzlibrary{calc} 
\usetikzlibrary{patterns}
\usepackage{algorithm}
\usepackage{algorithmicx}
\usepackage[noend]{algpseudocode}
\usepackage{rotating}
\usepackage{xspace}

\newcommand{\forbiddenZone}[4]{%
    \draw[red, thick, fill=red, fill opacity=0.1, pattern=checkerboard, pattern color=red] (#1, #2) rectangle (#3, #4);
}

\newcommand{\case}[3]{\medskip\noindent\textbf{\makebox[#3][l]{Case #1:} #2:}\par\nopagebreak}

\def\OSKP{{\scshape Oske}\xspace}

\def\N{{\bf N}}
\newcommand{\eps}{\varepsilon}

\newcommand{\notation}[2]{%
    \newcommand{#1}{\ensuremath{#2}\xspace}%
}

\notation{\kk}{k} 

\notation{\cri}{\overline{c}}
\notation{\currKP}{m} 

\notation{\tinyItem}{\mu} 
\notation{\midItem}{\nu} 

\usepackage{subfiles}

\begin{document}

\maketitle

\begin{abstract} 
Imagine you are a computer scientist who enjoys attending
conferences or workshops within the year. Sadly, your travel budget is limited,
so you must select a subset of events you can travel to.
When you are aware of all possible events and their costs
at the beginning of the year, you can select the subset of the possible
events that maximizes your happiness and is within your budget.
On the other hand, if you are blind about the options, you will likely have a hard time when 
trying to decide if you want to register somewhere or not, and will likely
regret decisions you made in the future. 
These scenarios can be modeled by knapsack variants, either by an offline 
or an online problem. However, both scenarios are somewhat unrealistic:
Usually, you will not know the exact costs of each workshop at the beginning of
the year. The online version, however, is too pessimistic, as you might already
know which options there are and how much they cost roughly.
At some point, you have to decide whether to register for some workshop, 
but then you are aware of the conference fee and the flight and hotel prices.

We model this problem within the setting of online knapsack problems with estimates:
in the beginning, you receive a list of potential items with their estimated size
as well as the accuracy of the estimates. 
Then, the items are revealed one by one in an online fashion with their actual size, and you need
to decide whether to take one or not.
In this article, we show a best-possible algorithm for each estimate accuracy $\delta$ 
(i.e., when each actual item size can deviate by $\pm \delta$ from the announced size) for 
both the simple knapsack and the simple knapsack with removability.
\end{abstract}

\newpage
\section{Introduction}

In the \textsc{Online (Simple) Knapsack} problem, items are presented one after another
to an algorithm, which then has to decide  on the spot whether to include the current item
into the knapsack of unit size, or not.
In the classical setting, it is easy to see that this problem
is non-competitive: A tiny item can be presented at the beginning and an algorithm
cannot know whether it should pack this item (as it might be the only one) or reject it 
(as there might be an item of size $1$ next, which the algorithm cannot pack when having packed
the tiny item). Therefore, the ratio between a solution of an online algorithm and an optimal (offline) solution on the same instance (the so-called \emph{competitive ratio}) can be arbitrarily high.

Arguably, one can say that this counterexample is rather pathological, as the restriction for deterministic, irrevocable decisions in a setting without any knowledge about the future is harsh.
So even though this (negative) example by Marchetti-Spaccamela and Versallis~\cite{MSV95} might seem demotivating for further research at first glance, it indeed has resulted in even more attention to online knapsack problems and their variants. For example, by allowing (limited) decisions to be revoked or delayed, or by relaxing the complete blindness of an online algorithm, many of those modified settings aim to avoid pathological counterexamples and therefore to model real-world knapsack applications more realistic. We will present some of those variants in the following section. In several cases, a variant studied for knapsack problems later turned out to be realistic for other online problems. For a more thorough
introduction to competitive analysis, we refer to the books by Borodin
and El-Yaniv~\cite{Borodin1998} and by Komm~\cite{Komm16}.

When acting in a real-world setting, the assumption of complete blindness is as unrealistic as the assumption of having correct information available: When navigating to some place, one usually is not aware of every detail (e.g., a traffic jam that might occur) but still has some information about the structure of the map and how long which routes roughly will take. The details then get revealed when arriving at some street, and seeing how the situation is. Or, when considering a packing problem such as bin packing or knapsack, one usually does not know the exact sizes of all items that are in the instance - but by experience, one might have a rough idea about, e.g., the items that need to be distributed into moving boxes for a moving - even though one likely does not want to measure all of them before packing to ensure that a packing is optimal in the end. Even if one have all the necessary information of some problem measured and, in theory, could compute an optimal solution, the exact values likely get lost when saving them on memory with limited size at the latest. 

In a sense, the setting of online problems with estimates lies in between offline problems with complete knowledge about the instance and classical online problems without any information about the future.

For the online (simple) knapsack problem, we assume an algorithm is given a list of items and their estimated size, together with a constant $\delta$ as the estimate accuracy. Each item with its actual size then gets revealed one after another, like in the classical online knapsack setting. Each item size can deviate by up to $\delta$ from their estimate.
An algorithm then has to decide about whether to pack an item or not: 

\newpage

In the case of irrevocable decisions, we show that no algorithm can achieve a competitive ratio better than $\frac1{\min(p,q)}$, with
\begin{equation*}
p = -
 \frac{0.5}{\lfloor \kk \rfloor} +\sqrt{\frac{1}{4\lfloor \kk \rfloor^2}+\frac{1-2\delta}{\lfloor \kk \rfloor}}
\text{ and } q = 1-2\delta-\frac{1}{\lceil \kk \rceil}\; \text{, where } \kk = \frac{2}{1-2\delta}.
\end{equation*}
for all $0 < \delta < \frac{1}{2}$ using three different instance constructions for different $\delta$. For a $\delta \geq \frac{1}{2}$, we show that the problem becomes uncompetitive. For $\delta < \frac{1}{2}$, where competitive algorithms can exist, we also present an algorithm that matches our lower bound and prove the tightness of of the behavior.

In the second part of the article, we consider the online simple knapsack problem with removability. Here an item, that was packed by an algorithm, can also be discarded at a later point. This very classical online knapsack variant avoids the pathological counterexample mentioned earlier: the tiny item can be packed without concerns first, and just be replaced by an item of size $1$ if one is presented in the instance. This setting was found to be $\Phi$-competitive by Iwama and Taketomi~\cite{IwamaT02}. 
In this case, we show that no algorithm can achieve a better competitive ratio than $\frac{3-2\delta}{2-2\delta}$ for all $0 <\delta < \frac{3}{4} - \sqrt{\frac54}$, and also present an algorithm that achieves this ratio as a tight upper bound. For all $\delta \geq \frac{3}{4} - \sqrt{\frac54}$, it turns out that the estimates will not help any longer, therefore the best possible algorithm is given by Iwama and Taketomi and achieves a competitive ratio of $\Phi$.

Finally, we conclude with some final remarks and open questions that arose while working
in this knapsack setting.

\subsection{Online Knapsack Problems}

 One way to deal with those artificial lower bounds is to allow an algorithm
 to fill a knapsack in a way a classical knapsack algorithm is not allowed to do:
A variant where the online algorithm is allowed to overpack the knapsack slightly is
called \emph{resource augmentation} by Iwama and
 Zhang~\cite{IwamaZ10}. Han and Makino~\cite{HanM10} allowed a limited number
 of \emph{cuts}, i.e.\ splitting of items into two sub-items.  
 
 Another approach is to delay decisions for some time.
We already mentioned a variant introduced by Iwama and Taketomi,
 in which packed items could be removed (also called \emph{preempted})
 from the knapsack, but not to be packed again \cite{IwamaT02}.
Another variant allows an algorithm to intermediately
 store items in a \emph{buffer} of a certain size, as introduced by Han
 et al.~\cite{HanKMY19}. The recently introduced
 model of \emph{reservation costs} allows an algorithm pay a fee to delay a decision for an arbitrary amount of time~\cite{BockenhauerBHLR21, burjons23, burjons2025}.
 
 Other variants allow the knapsack to grow over time, as in 
 Thielen, Tiedemann, and Westphal~\cite{ThielenTW16}, where a model in which the capacity of the knapsack increases step-wise over a given number of
 periods is studied.  
 
 It is also possible to avoid artificial instances by restricting the instance
 classes as in Zhou, Chakrabarty and Lukose~\cite{ZhouCL08}, where they analyzed the online knapsack
 problem under the assumption that the size of each item is much
 smaller than the knapsack capacity and the ratio between the value and
 the weight of an item is bounded within a given range.

 Further variations include randomization or
an oracle to communicate information about the instance
 via so-called \emph{advice bits} as introduced by
 Böckenhauer et al.~\cite{bockenhauer24, BockenhauerKKR14}. 

One of the main criticisms of the advice model is that the existence of 
an almighty oracle is not realistic in practice. One way to deal with this setting
is the model of \emph{machine learned advice}, which has recently been called
 \emph{untrusted predictions} or just \emph{predictions}. 
 Here an algorithm is usually given a prediction on each piece of the input,
 which tells the algorithm what to do with the piece of input.
 However, these predictions might be wrong and no bound is given on the error.
Therefore, the goal of an algorithm in this setting is to deal with those possibly
wrong hints, and ideally compute an optimal solution in case the predictions are
 correct (\emph{consistency}), but also performs as well as a regular
 online algorithm on the problem when the predictions become
 arbitrarily bad (\emph{robustness}) and ideally degrades with
 increasing unreliability of the prediction (\emph{smoothness}).

 The classical prediction model has seen a big influx of results in the past few
 years, with the model being applied to several different online
 problems, such as scheduling~\cite{LattanziLMV20, BoyarFKL23,
 	BalkanskiGT23}, metric algorithms~\cite{AntoniadisCE0S20,
 	AntoniadisCLPS23}, matching problems~\cite{DinitzILMV21,JinM22},
 spanning tree problems~\cite{ErlebachLMS22,BergBFL23}.
 
 Im et al.~\cite{ImKQP21} recently looked at the general knapsack problem, 
 which they studied under a model predicting the frequency of items of each size.
 Angelopoulos, Kamali, and Shadkami~\cite{AngelopoulosKS22} look at the online bin packing
 problem with predictions on the frequency of item sizes in the
 instance. Boyar, Favrholdt, and Larsen~\cite{BoyarFL22}  recently
 studied the online simple knapsack problem with predictions, but working with
 predictions on the \emph{average} size of the items an optimal
 solution would pack. Xu and Zhang~\cite{XuZ23} recently studied the simple
 knapsack problem in a learning-augmented setting, where they design
 algorithms that can learn and use the error of prediction.
 
Even though the prediction model feels aligned with the research in this work,
the focus lies on different aspects of inaccuracy: In the prediction setting,
the measure of accuracy is commonly defined by the number of correct and wrong
hints. In the setting of this article, we assume that the whole input might
not be as announced, but still lies in some surrounding of the announcement
(which size defines the accuracy).

 While we assume that an adversary can control both the
 predicted instance and the actual distortion of the items, there is a
 related model of \emph{smoothed analysis}, in which an adversary can
 fix an instance, which is then subject to some random (commonly
 Gaussian) distortion, or \emph{noise}. This model of an adversary without
 complete control over its prepared instance was first made
 popular when showing that the simplex algorithm runs in expected
 polynomial time when its input is subjected to such random
 noise~\cite{SpielmanT01}. Since then, there has been a large influx of
 results in this area for a wide range of problems, for example, the 0/1
 knapsack problem~\cite{BeierV03}. The model of smoothed analysis thus
 gives evidence that the worst-case running time or worst-case
 approximation ratios often seem to suffer from very specific and
 limited adversarial inputs which break down if even only a very slight
 perturbation of the instance is given - just as our pathological counterexample
 for the online simple knapsack.
 
 Furthermore, our model is related to robust optimization:
 The \emph{robust knapsack problem} by Monaci, Pferschy and Serafini~\cite{MonaciPS13}
 is very similar in that it also allows for an uncertain input with a multiplicative
 factor, but the authors look at \emph{offline} algorithms that see the
 complete permuted instance at once and are compared to the
 performance of a non-perturbed instance.  
 
\newpage
\subsection{Online Problems with Estimates}
The setting of online problems with estimates, where first a rough idea of the instance is given, and then the actual values gets revealed in an online manner, is rather new in online computation. 

Azar et al. considered a scheduling variant where the size of a job presented upon its revelation might not be given exactly~\cite{azar2021flow}. Here, they studied algorithms for the scenario where the accuracy of the estimate is known to the algorithm, as well as the scenario where this is not the case~\cite{azar2022distortion}. These results later got extended to a setting with multiple machines~\cite{azar2022distortion2}. While in these cases (as in our article) the estimates are given adversarially, Scully et al.~\cite{scully2021uniform} analyzed a setting where the actual durations of each job were picked randomly instead.
Azar et al. also speak of \emph{problems with predictions} within their works.

To avoid confusion with the previously mentioned model of predictions,
which usually refers to a setting where suggestions are given which might be wrong,
we name this specific type of prediction \emph{estimate} in the context of this work.

As packing problems are natural for a setting with estimated item sizes, we see that the knapsack setting with multiplicative accuracy behaves differently than the additive setting ~\cite{gehnen2024}: For example, while also not achieving a competitive ratio better than $2$ even for very small $\delta$, in the multiplicative setting an algorithm can achieve this ratio even up to a $\delta=\frac{1}{7}$. Furthermore, the divergence did not start at $\frac{1}{2}$ but at $1$. This different behavior is mainly caused by the fact that in the multiplicative setting, an adversary cannot present an item with the size of $0$ when the given estimate allowed also allowed to present the item as a non-zero item.

Apart from scheduling and packing problems, situations where the details are only revealed over time can also occur on graph problems. For the traveling salesman/graph exploration problem, it was shown that no algorithm can achieve a competitive ratio better than the estimate accuracy when the algorithm is aware of the graph structure with the estimated edge weights in the beginning, and the exact values are only revealed in a graph-exploration manner~\cite{gehnen2025}.
\subsection{Formal Definition}
To avoid confusion, we start with a formal definition of the problems we investigate and of the competitive ratio in the version that is referred to for the results. 
In this article, notation will be slightly abused by using $x_i$ for both the label of an item and the size of the same item, as its meaning is also clear in the given context.
\begin{definition}[The \textsc{Online Simple Knapsack with Item Size Estimates} Problem]\label{def:problem}
    Given an estimate accuracy $\delta \geq 0$, an instance of the \emph{Online Simple Knapsack with Item Size Estimate} consists of a series of \emph{items} $I = (x_1, \ldots, x_n)$, where each \emph{item} is a real number in $[0,1]$.
    At the beginning, the accuracy $\delta$ is announced as well as the \emph{item size estimates}  $P = (x_1', \ldots, x_n')$, such that for each $i \in \{1,\ldots,n\}$, $x_i' - \delta \le x_i \le x_i' + \delta$.  At each step $i \in \{1,\ldots,n\}$,
	the actual item size $x_i$ of the request sequence gets revealed. An algorithm then
	has the option to either pack the item in an initially empty knapsack $K$, if it fits or to reject the item:
	\begin{itemize}
		\item \makebox[3cm][l]{\textbf{Pack}}    If $\sum_{x_k \in K} x_k + x_i \leq 1$, set $K := K \cup x_i$\,.
		\item \makebox[3cm][l]{\textbf{Reject}}  Do nothing.
	\end{itemize}
	
	The \textsc{Online Simple Knapsack with Item Size Estimates} problem 
	is then for an algorithm $\mathtt{ALG}$ to minimize the competitive ratio between the size of his packing, compared to the optimal solution in the same instance.
\end{definition}

The online simple knapsack with removability and item size estimates can be defined analogously:
\begin{definition}[The \textsc{Online Simple Knapsack with Removability and Item Size Estimates} Problem]
    Given an estimate accuracy $\delta \geq 0$, an instance of the \emph{Online Simple Knapsack with Removability and Item Size Estimate} consists of a series of \emph{items} $I = (x_1, \ldots, x_n)$, where each \emph{item} is a real number in $[0,1]$.
    At the beginning, the accuracy $\delta$ is announced as well as the \emph{item size estimates}  $P = (x_1', \ldots, x_n')$, such that for each $i \in \{1,\ldots,n\}$, $x_i' - \delta \le x_i \le x_i' + \delta$.  At each step $i \in \{1,\ldots,n\}$,
	the actual item size $x_i$ of the request sequence gets revealed. An algorithm then
	has the option to remove a subset of items from the initially empty knapsack $K$, and then to either pack the current item into $K$, if it fits, or to reject the item:
	\begin{itemize}
        \item \makebox[3cm][l]{\textbf{Remove}}  Discard $S\subseteq K$ from the knapsack, set $K:= K-S$.
		\item \makebox[3cm][l]{\textbf{Pack}}    If $\sum_{x_k \in K} x_k + x_i \leq 1$, set $K := K \cup x_i$\,.
		\item \makebox[3cm][l]{\textbf{Reject}}  Do nothing.
	\end{itemize}
	
	The \textsc{Online Simple Knapsack with Removability and Item Size Estimates} problem 
	is then for an algorithm $\mathtt{ALG}$ to minimize the competitive ratio between the size of his packing, compared to the optimal solution in the same instance.
\end{definition}

Online algorithms are generally analyzed using competitive analysis, as introduced by Sleator and Tarjan~\cite{SleatorT84}. Simply speaking, it compares the solution quality of an online algorithm to the solution of an optimal offline algorithm in the same instance. As the knapsack problems are maximization problems, the following definition is suitable:
\begin{definition}
The \emph{(strict) competitive ratio} of an online algorithm $A$ is the highest ratio of any request sequence $S$ between
the gain of $A$ on $S$ and the gain of an algorithm $\mathtt{OPT}$ solving the problem optimally on the same request $S$,
\[{\rm CR}(\mathtt{A})=\sup_{ S }\left\{\frac{{\rm gain}_{\mathtt{OPT}}(S)}{{\rm gain}_{\mathtt{A}}(S)}\right\}\;.\]
\end{definition}
In other settings, a non-strict variant of the competitive ratio is commonly used as well, where an additional constant $k$ is allowed when comparing the solution quality of an online algorithm to the offline solution. Furthermore, a similar variant can also used to define the competitive ratio for minimization problems.

\section{Online Simple Knapsack}

In this section, we first show that the competitive ratio of the online simple knaspack problem with item size estimates is not less than $2$ for any $\delta >0$ and 
monotonously rises until it gets unbounded for any algorithm for
$\delta \geq 0.5$. 

Afterwards, we provide an algorithm that matches the competitive ratio of the lower bound.

We will see that the competitive ratio for all $0 < \delta < \frac{1}{2}$ is given by
$\frac1{\min(p,q)}$, where
\begin{equation}\label{eq:definition-of-p-and-q}
p = -
 \frac{0.5}{\lfloor \kk \rfloor} +\sqrt{\frac{1}{4\lfloor \kk \rfloor^2}+\frac{1-2\delta}{\lfloor \kk \rfloor}}
\text{ and } q = 1-2\delta-\frac{1}{\lceil \kk \rceil}\; \text{, where } \kk = \frac{2}{1-2\delta}.
\end{equation}

\begin{figure}
  \centering
  \begin{tikzpicture}
    \begin{axis}[ymin=0,
                 ymax=10,
                 xmin=0.01,
                 xmax=0.5,
                 x label style={at={(axis description cs:0.5,-0.1)},anchor=north},
                 xlabel=Distortion $\delta$,
                 ylabel=Competitive Ratio,
                 grid=major,
                 samples=100,
                 width=\textwidth,
                 height=7cm,
                 xticklabel style={/pgf/number format/.cd,fixed,precision=2}]
        \addplot[name path=ub2, domain=0:0.4] {max((1/(1-2*x-1/ceil(1/(0.5-x)))),(1/(-0.5*(1/floor(1/(0.5-x)))+sqrt(((1/floor(1/(0.5-x)))^2)/4+(1-2*x)/floor(1/(0.5-x))))))};


    \end{axis}
  \end{tikzpicture}
  \caption{Competitive ratio in the \emph{absolute error} model, depending on $\delta$}
  \label{fig:kp-fuzzy:add:results}
\end{figure}
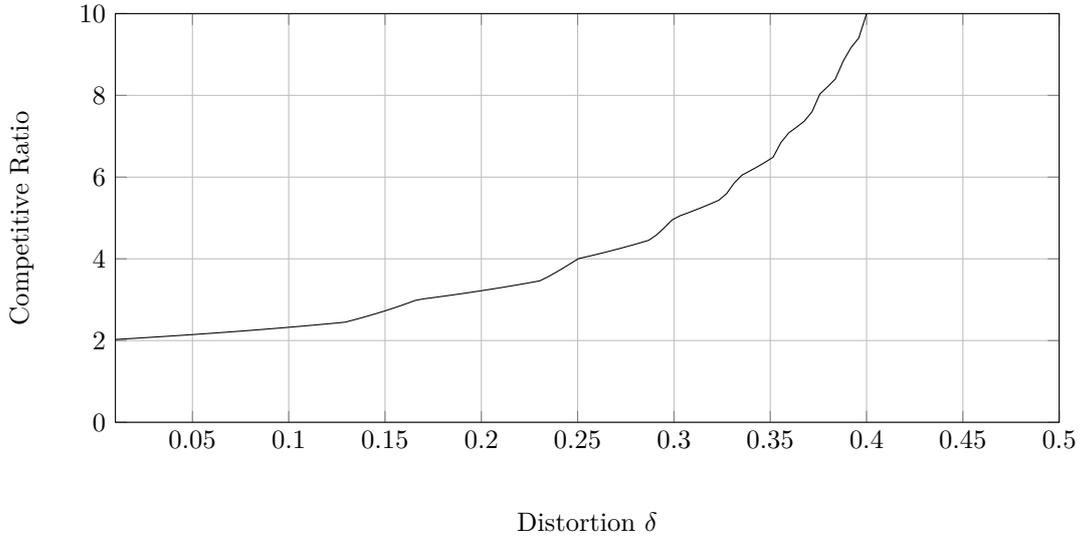

This ratio is visualized in Figure~\ref{fig:kp-fuzzy:add:results}.

Note that $p$ is the positive solution of the following equation.
\begin{equation}\label{eq:defining-equation-p}
\left\lfloor\frac{2}{1-2\delta}\right\rfloor \cdot \frac{p}{1-p-2\delta} = \frac{1}{p}
\end{equation}
The following inequalities, which hold for $0 \le \delta \le 0.5$, will turn out to be useful:
\begin{equation}\label{eq:pq-properties}
    1-p-2\delta \le p \text{ and } 1-q-2\delta = \left\lceil\frac{2}{1-2\delta}\right\rceil^{-1} \le q
\end{equation}

\subsection{Lower Bounds} \label{sec:kp-fuzzy:add:lower}

The following two lower bounds combined are best possible for all values of
$\delta$ up to $\frac{1}{2}$. Both functions have jumps induced by rounding effects and dominate
one another periodically. See Equation (\ref{eq:definition-of-p-and-q}) for the definition of $p$ and $q$. In the end, we note that for $\delta\geq \frac{1}{2}$, no constant competitive online algorithm can exist.

While the construction for the first bound is slightly more involved, it defines a
bound for the whole range of $\delta$ for $0<\delta<\frac{1}{2}$.
The second bound, as defined in Theorem~\ref{thm:kp-fuzzy:add:lower:qsmaller}, however only works starting from $\delta \geq \frac{1}{6}$. For values smaller than $\frac{1}{6}$, a different construction as in Theorem~\ref{lem:kp-fuzzy:add:lower:qsmaller:16} is necessary.
 
The crucial case in the following proof is
Case~{1.2}, where the knapsack of the optimal solution can be forced by a
clever algorithm not to be completely filled.

\begin{theorem}\label{thm:kp-fuzzy:add:lower:psmaller} For every $0 <
	\delta < 0.5$, 
 there exists no algorithm
	solving the \OSKP\ problem with a competitive ratio better than
	$\frac1p$.
\end{theorem}
\begin{proof}
	Let $0 < \varepsilon < \min(p,\delta)$ such that $1/\varepsilon \in \N$, and let $k = \left\lfloor\frac{2}{1-2\delta}\right\rfloor$.
Let us consider an arbitrary algorithm for the \OSKP\ problem.
The following instance is announce as a prediction:
	\[
	P = \left(
	\underbrace{
		\varepsilon,
	}_{1/\varepsilon \text{ many\;\; }}
	\underbrace{
		\frac{1}{2},
	}_{k \text{ many}}
	1 -p -\delta
	\right)
	\]
	We do a full case distinction on the possible behaviors of the
	algorithm. The first item is presented as $\varepsilon$.
	
	\case{1}{The algorithm packs \boldmath$\varepsilon$}{2.3cm}
	The subsequent $1/\varepsilon - 1$ items
	are presented as $0$. The next item is revealed with size $p$.
	This is possible as $p+\delta \geq 1/2$, which follows from Equation (3). 
	
	\case{1.1}{The algorithm packs \boldmath$p$}{2.3cm}
	All remaining items are presented as $1-p$, which the algorithm can not pack due to the $\varepsilon$-item.
    As we have seen, it is possible that an item which was announced of size $1/2$ can be presented as $p$;
    therefore presenting those items as $1-p$ is possible as well due to the symmetry of the deviation.
 
    An optimal solution can pack both, $p$ and its counterpart, while the algorithm has packed $p$ and an item of size $\varepsilon$.
    The competitive ratio is then $1/p$ for $\varepsilon$ converging to $0$.
	
	\case{1.2}{The algorithm rejects \boldmath$p$}{2.3cm}
	The subsequent $k -	1$ items will be presented as $p$ if the algorithm continues to reject them.
    If an algorithm should pack any of these items, we can use the same argumentation as in Case 1.1.
	
	Should the algorithm reject all $k$ items of size $p$, the last item is presented as $1-p-2\delta$.
    The optimal solution consists of all $k$ items of size $p$,
    whereas the algorithm only has the item of size $\varepsilon$ and the last item of size $1-2\delta-p$ in its knapsack.
    Since $1-p-2\delta \le p$ by Equation (\ref{eq:pq-properties}) and $kp / (1-p-2\delta) = 1/p$ by Equation (\ref{eq:defining-equation-p}) is valid,
    the competitive ratio is at least $1/p$ for $\varepsilon$ converging to $0$.        
    
	\case{2}{The algorithm rejects \boldmath$\varepsilon$}{2.3cm}
	The subsequent $1/\varepsilon - 1$ items will be presented as $\varepsilon$ if the algorithm continues to reject them.
    If the algorithm should pack any of these items, the remaining items are presented as $0$ and we use the same argumentation as in Case 1.
	
	Should the algorithm reject all $1/\varepsilon - 1$ items of size $\varepsilon$, the subsequent $k$ items will be presented as $p+\varepsilon$.
    If such an item is packed, the argumentation is the same as in case 1.1, just with the difference that the optimal solution contains the $\varepsilon$-items with a total size of $1$.
 
    If no such item is packed, the last item is presented of size $1-2\delta-p \le p$ (see Equation (\ref{eq:pq-properties})).
    The optimal solution packs the full knapsack using items of size $\varepsilon$.
    Again, the competitive ratio is at least $\frac1p$.
\end{proof}
The following theorem gives a better bound than Theorem \ref{thm:kp-fuzzy:add:lower:psmaller} when $q < p$ happens.
Note that $q > p$ is true for $\frac{1}{6} < \delta < \frac{1}{24}(9-2\sqrt{3}) \approx 0.23$ and the relation $\frac{3}{16} \in [\frac{1}{6},0.23]$ holds.
Consequently, after Theorem \ref{thm:kp-fuzzy:add:lower:qsmaller}, it remains to inspect the lower $q$-bound for $\delta < 1/6$.
\begin{theorem}\label{thm:kp-fuzzy:add:lower:qsmaller} For every
	$\frac{3}{16} < \delta < 0.5$, 
        there exists no
	algorithm solving the \OSKP\ problem with a competitive ratio better
	than $\frac1q$.
\end{theorem}
\begin{proof}
	Let $\varepsilon > 0$ such that $1/\varepsilon \in \N$, let $k = \left\lceil \frac{2}{1-2\delta} \right\rceil$, and recall that $q = 1 - 2\delta - 1/k$. Let us consider an arbitrary algorithm for the \OSKP\ problem.
The following prediction is announced to the algorithm:
	\[
	P = \left(
	\underbrace{
		\varepsilon,
	}_{1/\varepsilon \text{ many \;\;}}
	\underbrace{
		\delta,
	}_{k \text{ many \;\;}}
		q+\delta
	\right)
	\]
	
	\newpage
	We do a case distinction on the potential behaviors of the
	algorithm. The first item is presented as $\varepsilon$.
	
	\case{1}{The algorithm packs \boldmath$\varepsilon$}{2.3cm}
	The subsequent $1/\varepsilon - 1$ items are presented as $0$.
    The next presented item is of size $1/k$.
    This is possible if $2\delta \geq q$ since Equation (\ref{eq:pq-properties}) yields $q \geq 1/k$.
    The former holds for $\delta \geq 3/16$.
	
	\case{1.1}{The algorithm packs \boldmath$1/k$}{2.3cm}
    The remaining $k - 1$ items are presented as $0$, with the last item presented as $1 - 1/k = q + 2\delta$.
    An optimal solution can pack both $1/k$ and its counterpart, while the algorithm has packed $1/k$ and an item of size $\varepsilon$.
    The competitive ratio is then $k$ for $\varepsilon$ converging to $0$ which yields the wished ratio by $k \geq 1/q$ tanks to Equation (\ref{eq:pq-properties}).
	
	\case{1.2}{The algorithm rejects \boldmath$1/k$}{2.3cm}
	The subsequent $k - 1$ items will be presented as $1/k$, with the last item presented as $q$, if the algorithm continues to reject them.
    If an algorithm should pack any of these items, we can use the same argumentation as in Case 1.1.
	
	The optimal solution consists of all $k$ items of size $1/k$, which add up to exactly $1$,
    whereas the algorithm only has the item of size $\varepsilon$ and the last item $q$ in its knapsack.
    The competitive ratio is again at least $1/q$ for $\varepsilon$ converging to $0$.

    \case{2}{The algorithm rejects \boldmath$\varepsilon$}{2.3cm}
    The subsequent $1/\varepsilon - 1$ items will be presented as $\varepsilon$ as long as they get rejected.
    If an algorithm packs any of these items, we can use the same argumentation as in Case 1.

    Should the algorithm reject all $1/\varepsilon - 1$ items of size $\varepsilon$, the subsequent $k$ items will be presented as $0$.
    The last item is presented as $q$ which directly yields the competitive ratio of $1/q$.
    Here an optimal solution consists of a full knapsack with $\varepsilon$ items.
    \end{proof}

The crucial issue of Theorem~\ref{thm:kp-fuzzy:add:lower:qsmaller} for small $\delta$ are the items of announced size $\delta$, where an adversary must be able to present them as a $0$ or as $\frac{1}{3}$ for all $\delta<\frac{1}{6}$. As this is not possible, we need to handle the setting $\delta < \frac{1}{6}$ as a special case.
In particular, $q > p$ for $0 < \delta < \frac{1}{12}(4-\sqrt{6}) \approx 0.129$ is true, and since $1/12 < 0.129$ holds, starting with delta at $1/12$ is no limitation.

\begin{theorem}\label{lem:kp-fuzzy:add:lower:qsmaller:16}
    For every $\frac{1}{12} < \delta < \frac{1}{6}$,
    there exists no algorithm solving the \OSKP\ problem with a competitive ratio better than $\frac1q$.
\end{theorem}

\begin{proof}
    We define $a = \frac{1}{3}-2\delta$, and let $\varepsilon > 0$ be arbitrary such that $1/\varepsilon \in \N$ and $a/\varepsilon \in \N$ holds.

    Let us consider an arbitrary algorithm for the \OSKP\ problem.
    The algorithm receives the following prediction:
    \[
	P = \left(
	\underbrace{
		\varepsilon,
	}_{1/\varepsilon \text{ many \;\;}}
	\underbrace{
		\frac{1}{3}+\delta
	}_{3 \text{ many \;\;}}
	\right)
	\]

    We do a similar full case distinction on the possible behaviors of the algorithm, like in Theorem \ref{thm:kp-fuzzy:add:lower:psmaller} and Theorem \ref{thm:kp-fuzzy:add:lower:qsmaller}.
    The first item is presented as $\varepsilon$, but this time we stop presenting $\varepsilon$-items when the algorithm has packed exactly an amount of $y := a + \epsilon$.

	\newpage
	\case{1}{The algorithm packs an amount of \boldmath$y = a + \varepsilon$ with \boldmath$\varepsilon$-items}{2.3cm}
	The subsequent $\varepsilon$-items are presented as $0$.
    The next item is presented of size $1/3$.
	
	\case{1.1}{The algorithm packs \boldmath$1/3$}{2.3cm}
    The remaining $2$ items are presented as $\frac{2}{3} - a = \frac{1}{3} + 2\delta$.
    An optimal solution can pack $1/3$ and its counterpart, while the algorithm has packed $1/3$ and some $\varepsilon$-items of total size $a+\varepsilon$.
    The competitive ratio is then $(2/3 - 2\delta)^{-1} = (1-1/3 - 2\delta)^{-1} = 1/q$ for $\varepsilon$ converging to $0$ which yields the wished competitive ratio.
	
	\case{1.2}{The algorithm rejects \boldmath$1/3$}{2.3cm}
	The subsequent $2$ items will be presented as $1/3$ until one gets accepted.
    If an algorithm accepts one of the first two items, we can use the same argumentation as in Case 1.1.

    Otherwise, the algorithm accepts the last $1/3$-item or none of them.
	The optimal solution consists of all $3$ items of size $1/3$, which add up to exactly $1$,
    whereas the algorithm again has only the $\varepsilon$-items of total size $a+\varepsilon$ and at most the last $1/3$ item in its knapsack.
    The calculation of the competitive ratio is analogous to case 1.1.

    \case{2}{The algorithm do not pack at least \boldmath$y = a + \varepsilon$ with $\varepsilon$-items}{2.3cm}
    We presented all the $\varepsilon$-items as $\varepsilon$ and we know after these items that we got $y \leq a < a + \varepsilon$ in our knapsack.

    Next, we start to reveal $1/3+(a-y)+\varepsilon$ items. This is possible because of the relation 
    $$\frac{1}{3}+\varepsilon \leq \frac{1}{3}+(a-y)+\varepsilon \leq \frac{1}{3}+a+\varepsilon = \frac{2}{3} - 2\delta + \varepsilon \leq \frac{1}{3} + 2 \delta \quad \text{for } \delta > \frac{1}{12}.$$
    Should the algorithm pick one of these $3$ items, we are in the same setting as in case 1.1, but with an optimal solution consisting of just $\varepsilon$-items.
\end{proof}

As the construction of Theorem \ref{thm:kp-fuzzy:add:lower:qsmaller} cannot work for small $\delta$, it is also worth noting that the construction of Theorem \ref{lem:kp-fuzzy:add:lower:qsmaller:16} is not working for larger $\delta$. This is mainly because the items which are announced with size $\frac{1}{3+\delta}$ must be presented of size $\frac{1}{3}$ at least. Even if the $3$ would be replaced with some $k$, it would allow an algorithm to pack multiple of those items without a chance of an adversary to hinder him.

Finally, we see that the case $\delta \ge 0.5$ can be handled using a standard construction by Marchetti{-}Spaccamela and Vercellis~\cite{MSV95}, thus no algorithm can achieve a bounded competitive ratio here.

\begin{theorem}\label{thm:kp-fuzzy:add:lower:noncomp} For every $\delta \ge 0.5$, there exists no algorithm solving the \OSKP\ problem with a constant competitive ratio.
\end{theorem}
\begin{proof}
Consider an arbitrary algorithm and the prediction $P = (0.5, 0.5)$. The first item arrives as $\varepsilon > 0$. If the algorithm rejects it, the second item arrives as 0. Otherwise, the second item arrives as $1 -\varepsilon/2$ so the algorithm cannot pack it, whereas the optimum solution can. This construction shows that no algorithm can be competitive.
\end{proof}

\subsection{Upper Bounds} \label{sec:kp-fuzzy:add:upper}
We start this section with a simple algorithm, that either takes the largest announced item or packs the knapsack in a greedy way.

\begin{algorithm}
	\caption{$\frac{2}{1-2\delta}$-competitive Algorithm for $0 < \delta < \frac12$.}
	\begin{algorithmic}
		\If{$b$ is the largest announced item and $b \geq 0.5$}
		\State Pack only $b$. END
		\Else
		\State Greedily pack items. END
		\EndIf
	\end{algorithmic}
	\label{alg:kp-fuzzy:add:upper:greedy}
\end{algorithm}
It turns out that this algorithm already matches the lower bound for all accuracies $\delta$ of the form $\frac12-\frac{1}k$ for integers $k \geq 3$.
\newpage
\begin{theorem}\label{thm:kp-fuzzy:add:upper:greedy}
	Given a fixed $\delta$ with $0 < \delta < \frac12$,
	Algorithm~\ref{alg:kp-fuzzy:add:upper:greedy} solves the \OSKP\ problem
	with a competitive ratio of at most $\frac{2}{1-2\delta}$.
\end{theorem}

\begin{proof}
	Assume that there exists an announced item of size at least
	$0.5$. Then the algorithm waits for it, packs it, and
	achieves a gain of $0.5-\delta$, which gives us the claimed bound.
	
	Thus, assume that such an item does not exists but that there exists
	an item that the algorithm cannot
	fit into its knapsack when packing greedily. This item can be at most
	of actual size $0.5+\delta$, meaning our gap due to not packing
	this item is at most $1-(0.5+\delta) = 0.5 -\delta$, which again gives
	us our wanted bound.
\end{proof}

In the rest of this section, we will present a more refined algorithm and prove that it matches the lower bounds of Theorems~\ref{thm:kp-fuzzy:add:lower:psmaller} and \ref{thm:kp-fuzzy:add:lower:qsmaller}.
Let us fix an instance $(P, \delta)$, where $0 < \delta < 0.5$ holds, and $P = (x_1', \ldots, x_n')$ are the announced item sizes of the \OSKP\ problem.

As in Definition~\ref{def:problem}, if $x$ is an item, then $x'$ denotes its announced size.

Let $c = 1/\min(p,q)$ and $\cri = 1/c$, see Equation (\ref{eq:definition-of-p-and-q}) for the definition of $p$ and $q$.

\begin{algorithm}
	\caption{$c$-competitive algorithm for $0< \delta < 0.5$.}
	\begin{algorithmic}[1]
    \If{there is an item $x$ such that $x' \geq \cri+\delta$}
        \State Pack only $x$. END.\EndIf
    \If{all items have announced size $\leq 1-\cri-\delta$}
        \State Pack greedily. END.\EndIf
    \State Let $x_l$ be the last item such that $x_l' \in (1-\cri-\delta, \cri+\delta)$.
    \State \textsf{greedy} := \textbf{false}.
    \While{there is a next item $y$ in the queue}
            \If{$y = x_l$} \State \textsf{greedy} := \textbf{true}\EndIf
            \State Let \currKP be the size of already packed items.
            \If{\textsf{greedy}} \State Pack $y$ if it fits into the knapsack.
            \ElsIf{$m \in [\cri - (x_l' - \delta),  1 - (x_l' + \delta)]$ or $y + \currKP \in (1 - (x_l' + \delta), \cri)$}

            \State Skip $y$.
            \Else \State Pack $y$ if it fits into the knapsack.
            \EndIf

    \EndWhile
    
\end{algorithmic}
	\label{alg:kp-fuzzy:add:upper:complex}
\end{algorithm}
It is easy to see that the algorithm achieves the desired competitive ratio in many instances.

\begin{lemma}\label{lem:add:upper:trivial-cases}
If the largest announced size is not in $( 1-\cri-\delta,\cri+\delta)$, then Algorithm \ref{alg:kp-fuzzy:add:upper:complex} achieves the competitive ratio of at least $\cri$.
\end{lemma}
\begin{proof}
First, suppose there is an item $x$ of announced size at least $\cri+\delta$.
In this case, $x$ is the only item packed, see line~2.
Since $x \ge \cri$ is true, the competitive ratio is ensured.

Second, suppose that all items have announced size at most $1-\cri-\delta$.
In this case, we pack greedily, see line~4.
Let $x$ be the first item that cannot be packed by the algorithm; if $x$ does not exist, we have found the optimal solution.
Otherwise, $x \le 1-\cri$ holds, which means that we have packed more than $\cri$ and the competitive ratio is also ensured.
\end{proof}

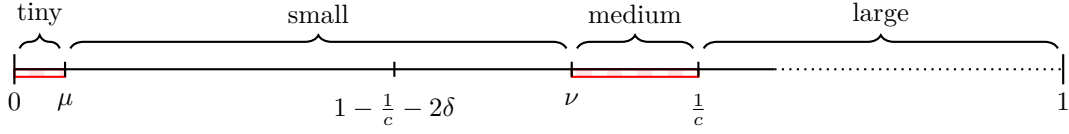
\begin{figure}
  \centering
  \begin{tikzpicture}
    \pgfmathsetmacro{\totalwidth}{13.8}
    \pgfmathsetmacro{\width}{10}
    \pgfmathsetmacro{\widthfactor}{1/0.3}
    \pgfmathsetmacro{\markheight}{0.1}
    \pgfmathsetmacro{\itemMu}{0.02}
    \pgfmathsetmacro{\itemQuarter}{0.15} 
    \pgfmathsetmacro{\itemNu}{0.22} 
    \pgfmathsetmacro{\itemC}{0.27} 
    \pgfmathsetmacro{\itemMu}{\itemMu*\width*\widthfactor}
    \pgfmathsetmacro{\itemQuarter}{\itemQuarter*\width*\widthfactor}
    \pgfmathsetmacro{\itemNu}{\itemNu*\width*\widthfactor}
    \pgfmathsetmacro{\itemC}{\itemC*\width*\widthfactor}

    \forbiddenZone{0}{0}{\itemMu}{-\markheight}
    \forbiddenZone{\itemNu}{0}{\itemC}{-\markheight}
  
    \draw[thick] (0,0) -- (\width,0); 
    \draw[thick,dotted] (\width,0) -- (\totalwidth,0); 
    \draw[thick] (0,2*\markheight) -- (0,-2*\markheight); 
    \draw[thick] (\totalwidth,2*\markheight) -- (\totalwidth,-2*\markheight);

    \draw[thick] (\itemMu,\markheight) -- (\itemMu,-\markheight);
    \draw[thick] (\itemQuarter,\markheight) -- (\itemQuarter,-\markheight);
    \draw[thick] (\itemNu,\markheight) -- (\itemNu,-\markheight);
    \draw[thick] (\itemC,\markheight) -- (\itemC,-\markheight);

    \node[below] at (0,-2*\markheight){0};
    \node[below] at (\totalwidth,-2*\markheight){1};
    \node[below] at (\itemMu,-2*\markheight){\tinyItem};
    \node[below] at (\itemNu,-2*\markheight){\midItem};
    \node[below] at (\itemQuarter,-2*\markheight){$1-\frac{1}{c}-2\delta$};
    \node[below] at (\itemC,-2*\markheight){$\frac{1}{c}$};

    \pgfmathsetmacro{\bracespace}{0.07}
    \draw[thick,decorate,decoration={brace,amplitude=5pt,raise=5pt},yshift=0pt]
        ($(0+\bracespace,\markheight)$) -- ($(\itemMu-\bracespace,\markheight)$) node [black,midway,yshift=18pt] {tiny};
    \draw[thick,decorate,decoration={brace,amplitude=5pt,raise=5pt},yshift=0pt]
        ($(\itemMu+\bracespace,\markheight)$) -- ($(\itemNu-\bracespace,\markheight)$) node [black,midway,yshift=18pt] {small};
    \draw[thick,decorate,decoration={brace,amplitude=5pt,raise=5pt},yshift=0pt]
        ($(\itemNu+\bracespace,\markheight)$) -- ($(\itemC-\bracespace,\markheight)$) node [black,midway,yshift=18pt] {medium};
    \draw[thick,decorate,decoration={brace,amplitude=5pt,raise=5pt},yshift=0pt]
        ($(\itemC+\bracespace,\markheight)$) -- ($(\totalwidth-\bracespace,\markheight)$) node [black,midway,yshift=18pt] {large};

  \end{tikzpicture}
  \caption{Range of item sizes. \textit{Forbidden} ranges of item sizes to be packed are marked in red.}
  \label{fig:item:sizes}
\end{figure}

For the rest of the section, suppose that the largest announced item is in $(1-\cri-\delta, \cri+\delta)$, 
and let $x_l$ be the last announced item such that $x'_l \in(1-\cri-\delta, \cri+\delta)$.
In this case, the algorithm can guarantee a packing of at least $\cri$ for most instances. To see this, we classify each item $y$ as either \emph{tiny, small, medium} or \emph {large} depending on the actual size of $y$ and the sum of the already packed items when $y$ arrives.
We use the substitutions $\tinyItem := \cri - (x_l' - \delta)$ and $\midItem := 1 - (x_l' + \delta)$.
To give an overview of the variables and relations we provide Figure \ref{fig:item:sizes}.

\begin{definition}\label{def:size-classification}
Let $y$ be an item and let $m$ be the sum of the already packed items when $y$ arrives. 
We say that an item $y$ is \emph{tiny} if $y+m \in [0, \tinyItem)$, \emph{small} if $y+m \in [\tinyItem,\midItem]$, \emph{medium} if $y+m \in (\midItem, \cri)$, and \emph{large} otherwise.

\end{definition}

Observe that before $x_l$ arrives, medium items are skipped by the algorithm, see line 13.
The case when a small or large item arrives is handled by the following lemma.

\begin{lemma}\label{lem:small-and-large}
If a small or large item $y$ arrives before $x_l$, 
then Algorithm \ref{alg:kp-fuzzy:add:upper:complex} reaches a packing of $\cri$.
\end{lemma}
\begin{proof}
Let $y$ be the first small or large item that arrives and let $m$ be the size of the packing when $y$ arrives. Since medium items are skipped, only tiny items were packed before $y$, which means $m < \tinyItem$.

First, suppose that $y$ is small.
In this case, $y$ is packed and after that, the knapsacks packing is of size in $[\tinyItem,\midItem]$, which means that no item is packed until $x_l$ arrives because the condition on line 13 is satisfied.
Observe that when $x_l$ arrives, the algorithm tries to pack it.
Note that $m + y + x_l \le m + (\midItem-m) + (x'_l + \delta) = 1$, which means that $x_l$ fits into the knapsack when it arrives.
Now observe that after $x_l$ is packed, the knapsack has size at least $m +y+x_l \geq m + ( \tinyItem-m) + (x'_l- \delta) = \cri$, as required.

Second, suppose that $y$ is large.
Note that $y < \cri + \delta$ since otherwise we would be in the case handled by Lemma~\ref{lem:add:upper:trivial-cases}.

Now observe that $m + y \le \tinyItem + (\cri+2\delta) =
2\cri - x_l' + 3\delta \le
2\cri - (1-\delta-\cri) + 3\delta = 3\cri +4\delta -1 \le 3q + 4\delta - 1$, where the last inequality follows from $\cri = \min(p, q)$.
Now we expand the definition of $q$, see Equation (\ref{eq:definition-of-p-and-q}): 
\[3q +4\delta -1 = 4\delta - 1 +3(1-2\delta - \frac{1}{\lceil \frac{2}{1-2\delta} \rceil}) = 2-2\delta - \frac{3}{\lceil \frac{2}{1-2\delta} \rceil} \le 2-2\delta - (1 -2\delta) = 1 \]
Hence, $y$ fits into the knapsack.
By Definition \ref{def:size-classification}, $y+m \ge \cri$ is valid, as required.

In both cases, the algorithm packs $y$ and can guarantee a packing of at least $\cri$.
\end{proof}

\newpage

Recall that the instance of \OSKP we are solving is $(P, \delta)$.
The following lemma says that we can assume that $x_l$ is the last item.

\begin{lemma}\label{lem:after-xl}
Let $P'$ be the prefix of $P$ ending with $x_l$ and suppose that Algorithm \ref{alg:kp-fuzzy:add:upper:complex} achieves a competitive ratio of at most $c$ on $(P', \delta)$.
Then, it also achieves a competitive ratio of at most $c$ on $(P, \delta)$. 
\end{lemma}
\begin{proof}
If all items after $x_l$ are packed, then the competitive ratio can only decrease.
Suppose that an item $x$ comes after $x_l$ and is not packed.
By choice of $x_l$, we know that $x' \le 1-\cri-\delta$, which means that $x \le 1 - \cri$.
Hence, the algorithm has packed more than $\cri$, and the competitive ratio is at most $c$.
\end{proof}

By Lemma \ref{lem:small-and-large} and \ref{lem:after-xl}, we may assume that $x_l$ is the last item and all other items are either tiny or medium.
We split the analysis into two cases depending on whether at most $\lfloor \frac{2}{1-2\delta} \rfloor$ non-tiny items are packed in an optimal solution.

\begin{lemma}\label{thm:upperboundp}
If no items are small or large, $x_l$ is the last item. Then there is an optimal solution packing 
at most $k := \left\lfloor \frac{2}{1-2\delta} \right\rfloor$ non-tiny items, then Algorithm \ref{alg:kp-fuzzy:add:upper:complex} achieves a competitive ratio of at most $\frac{1}{p}$.
\end{lemma}
\begin{proof}
Recall that the algorithm packs exactly the tiny items and $x_l$, which means we may assume that there are no tiny items (their presence would decrease the competitive ratio).
Since $x_l \ge 1-\cri-2\delta$ holds due to the choice of $x_l$, we may assume $x_l = 1-\cri-2\delta$ without decreasing the ratio.
By Equation (\ref{eq:pq-properties}), $1-\cri-2\delta \le \cri$ is valid, since $\cri = \min(p, q)$ is defined.
Hence, each item is of size at most $\cri$ and the ratio is at most $\frac{k \cri}{1-\cri-2\delta}$. By Equation (\ref{eq:defining-equation-p}), we have
\begin{align*}
    \frac{k \cri}{1-\cri-2\delta} \le \frac{k p}{1- p -2\delta} = \frac{1}{p}.
\end{align*}
Hence, the algorithm achieves a competitive ratio of $\frac{1}{p}$. This matches the lower bound given by Theorem \ref{thm:kp-fuzzy:add:lower:psmaller} for the case $p\leq q$.
\end{proof}
For the other case, we can guarantee that the knapsack is filled with at least $\frac{1}{q}$.
\begin{lemma}\label{thm:upperboundq}
If no items are small or large, $x_l$ is the last item. Then there is an optimal solution packing 
at least $k := \left\lceil \frac{2}{1-2\delta} \right\rceil$ non-tiny items, then Algorithm \ref{alg:kp-fuzzy:add:upper:complex} achieves a competitive ratio of at most $\frac{1}{q}$.
\end{lemma}
\begin{proof}
Since at least $k$ non-tiny items are packed by an optimal solution, there is a non-tiny item of size at most $1/k$. 

First, suppose that there is a medium item $y \ne x_l$ such that $y \le 1/k$. 
Let $m$ be the size of the packed items immediately before $y$ arrives.
Since $y$ is medium, we have $1/k + m \ge y + m > \midItem$.
Hence, after $x_l$ is packed, the algorithm has packed
$$  m + x_l \ge (\midItem -1/k) + (x_l' - \delta) = 1 - (x'_l + \delta) - 1/k + x_l' - \delta = 1 - 2\delta -1/k = q,   $$
which implies a competitive ratio of at most $1/q$.

The remaining case is $y > 1/k$ for each medium item $y \ne x_l$ and $x_l \le 1/k$.
Observe that
$$  x_l' \le 1/k + \delta = 1-(1-2\delta-1/k)-\delta = 1-q-\delta \le 1- \cri - \delta, $$
which is a contradiction since $x_l' > 1- \cri - \delta$.
Hence, this case cannot occur, and the algorithm achieves a competitive ratio of at most $\frac{1}{q}$.
\end{proof}

With the previous observations, the following theorem immediately follows.
\begin{theorem}\label{thm:kp-fuzzy:mult:upper:complex}
For every $\delta \in (0,\frac{1}{2})$, Algorithm \ref{alg:kp-fuzzy:add:upper:complex} achieves a competitive ratio of $\frac{1}{\min(p,q)}$.
\end{theorem}

\section{Online Simple Knapsack with Removability}

In this section, the algorithm is allowed to discard previously packed items.
We consider not only additive estimate accuracy but will later also note that multiplicative accuracy behaves very similarly.

\subsection{Lower Bounds}

\begin{theorem}\label{thm:lbremoval}
For every
$\delta \in (0, \frac{3}{4} - \sqrt{\frac54}] $, there exists no algorithm solving the \OSKP\
problem with removability and item size estimate with a competitive ratio better than $1/x$, where

\[
x = \frac{2-2\delta}{3-2\delta}
\]

\end{theorem}
\begin{proof}
Let $\eps > 0$ be small enough so that $x+2\eps \le x + \delta$ and $1-x-\eps \ge 1-x+\eps-\delta$.
An arbitrary algorithm is given the following prediction: \[\left(1-x, x + \eps, x, 1-x+\eps+\delta\right).\]
The first two items presented are $1-x$ and $x+\eps$. The algorithm can pack either of these items but not both.

\case{1}{The algorithm packs \boldmath $x + \eps$}{2.3cm}
The third item is presented as $x$ and the last one as $1-x+\eps$.
Since $x > 0.5$, the algorithm can pack at most one item from the set $\{x, x+\eps, 1-x+\eps\}$ and its best choice is keeping $x+\eps$, whereas an optimal solution is to pack $x$ and $1-x$. Hence, the competitive ratio goes to $1/x$ as $\eps$ goes to 0.

\case{2}{The algorithm packs \boldmath $1-x$}{2.3cm}
The next item is presented as $x+2\eps$. Now the algorithm either keeps $1-x$ or discards it and packs $x+2\eps$.

\case{2.1}{The algorithm packs \boldmath $x+2\eps$}{2.3cm}
The final item is presented as $1-x-\eps$.
Since $x+2\eps \ge 1-x-\eps$, the algorithm can pack at most $x+2\eps$, whereas the optimal solution packs the full knapsack by taking $x+\eps$ and $1-x-\eps$. Hence, the competitive ratio goes to $1/x$ as $\eps$ goes to 0.

\case{2.2}{The algorithm keeps \boldmath $1-x$}{2.3cm}
The final item is presented as small as possible and presented as $y := 1-x-2\delta+\varepsilon$. Now the algorithm can pack at most $1-x+y$ whereas the optimal solution packs $x + 2\eps + y$.
Hence, the competitive ration goes to $(1-x+y)/(x+y)$ as $\eps$ goes to 0.
By definition of $x$, we see that also, in this case, the algorithm cannot achieve a better competitive ratio than $\frac{1}{x} = c$.
\end{proof}

Note that the construction of Theorem~\ref{thm:lbremoval} will yield a bound of $\Phi$ for $\delta = \frac{3}{4} - \sqrt{\frac54}$, which can also be achieved without estimates as in Iwama and Taketomi~\cite{IwamaT02}.
\begin{theorem}
    For all $ \delta > \frac{3}{4} - \sqrt{\frac54} $, no algorithm for the online simple knapsack problem with removability and estimates can achieve a competitive ratio of less than $\Phi$.
\end{theorem}

\subsection{Upper Bounds}

In this section, we will present an algorithm that matches the lower bound of the previous part.
Again, define $x= \frac{2-2\delta}{3-2\delta}$.
We call items of size at most $1-x$ \emph{small} items, items of size strictly between $1-x$ and $x$ \emph{medium} items, and items of size at least $x$ \emph{large} items. 

\begin{algorithm}
    \caption{for Online Knapsack with Removability and Estimates}
    \begin{algorithmic}[1]
        \State Let $x_l$ be the last item such that $x_l' > 1-x-\delta$.
        \While{there is an item $y$ in the queue}
        \State \textbf{if} at least $x$ has been packed \textbf{then} END.
        \State \textbf{if} $y$ is large \textbf{then} Remove everything. Pack $y$. END.
        \State \textbf{if} $y$ is small \textbf{then} Pack $y$ if it fits.
        \If{$y$ is medium}        
        \If{no medium item is packed} \State Pack* $y$. 
        \Else
        \State Let $z$ be the medium item packed.
        \State \textbf{if} $y + z \le 1$ \textbf{then} Remove everything but $z$. Pack $y$. END.
        \State \textbf{else if} $y < z$ or $y = x_l > z$ \textbf{then} Remove $z$. Pack* $y$.
        \State \textbf{else} Ignore $y$.
        \EndIf
        \EndIf
        \EndWhile
    \end{algorithmic}
* If it is not possible to pack $y$ on lines 8 or 12, we keep removing small items until it becomes possible.\label{alg:removability}
\end{algorithm}

Note that once two medium items are packed, Algorithm~\ref{alg:removability} terminates, see line 11. 
Hence, line 10 is well defined as at most one medium item can be packed at that point.

Now we prove that Algorithm~\ref{alg:removability} matches the lower bound of Theorem~\ref{thm:lbremoval}.

\begin{theorem}
    Algorithm~\ref{alg:removability} achieves a competitive ratio at least $1/x$ for $\delta \leq \frac{3}{4} - \sqrt{\frac54}$.
\end{theorem}
\begin{proof}
    First, suppose that $y$ is the first large item in the instance.
    Observe that the algorithm packs $y$ on line 4 and terminates. Since $y \ge x$, a competitive ratio of at least $1/x$ is achieved.
    From now on, suppose that there are no large items.

    Second, suppose that there are two medium items $y_1$ and $y_2$ such that $y_1 + y_2 \le 1$.
    Observe that until $x_l$ arrives, if there is only one medium item packed, then it is the smallest medium item that has arrived so far (see line 12).
    Hence, there is an iteration of the while loop in which the condition on line 11 is satisfied, i.e., two medium items are packed and the algorithm terminates.
    Since $2(1-x) > x$ holds for $\delta > 0$, the algorithm has again packed at least $x$ as required.
    From now on, suppose that no two medium items fit together, no matter their position in the instance.

    Third, suppose there is a small item $a$ that is not part of the knapsack at the end, i.e., $a$ was ignored on line 5 or removed on line 8 or 12 before packing $y$.  
    Let $m$ be the value in the knapsack at the end of the iteration in which $a$ is discarded, and observe that $m + a > 1$ (otherwise $a$ would be packed, resp. not removed).
    Since $a \le 1 -x$, we have $m > 1- a \ge 1-(1-x) = x$.
    Hence, the algorithm has packed at least $x$ (and terminates in the next iteration, see line 3).
    From now on, we may assume that the algorithm packs all small items. 
    In particular, if there is at most one medium item, the algorithm finds the optimal solution.
    Hence, assume that there are at least two medium items.

\newpage
    Observe that by definition of $x_l$, there are no medium items after $x_l$.
    Hence, when $x_l$ arrives, there is a medium item $y$ in the knapsack.
    Let $s$ be the size of all small items except for $x_l$ (note that $x_l$ is small or medium).
    Assume $x_l$ is small, i.e., the algorithm packs at least $x_l + (1-x) + s$.
    Since no two medium items fit together, an optimal solution packs at most $x_l + x + s$.
    Observe that $x_l \ge 1-x-\delta -\delta$, as it was announced with size at least $1-x-\delta$.
    Hence the competitive ratio is at most:
    \[\frac{x_l + x+ s}{x_l+(1-x) + s} \le \frac{x + 1-x-2\delta}{1-x + 1-x-2\delta} = \frac{1}{x} \text{ as desired.}\]
Finally, suppose that $x_l$ is medium. 
Observe that the algorithm packs $\max(y, x_l) + s > 0.5 + s$, as $y$ and $x_l$ combined exceed $1$.
Since no two medium items fit together, an optimal solution packs at most $x + s$.
Hence, the competitive ratio is at most $x/0.5 < 1/x$ as desired.
\end{proof}

Again, if $\delta \leq \frac{3}{4} - \sqrt{\frac54}$, an optimal algorithm does not need to
consider the estimates given.
\begin{theorem}
For all $\delta >0$, there is an algorithm that achieves a competitive ratio of $\Phi$ for the online simple knapsack problem with removability and estimates~\cite{IwamaT02}.
\end{theorem}
\subsection{Multiplicative Estimate Accuracy}

While the previous research was dedicated to additive estimate accuracy, it turned out that the results easily translate to the multiplicative model (as studied in~\cite{gehnen2024}) with removability. This is not surprising, as the estimate accuracy is only needed for one item at the end in both the lower and the upper bounds, which has a medium size.

Therefore, it is possible to use an analogous lower bound construction and upper bound algorithm for the multiplicative case when the aimed competitive ratio is changed accordingly. In the multiplicative setting, $x$, which is defined as $\frac{2-2\delta}{3-2\delta}$ in the additive setting, can be redefined to $x = \frac{\sqrt{\delta^2+10\delta+9}+\delta-3}{4\delta}$, achiving a competitive ratio of $\frac{1}{x}$ then. As in the additive case, both bounds are tight until they reach $\Phi$ where the estimates become worthless due to the bound of $\Phi$ without estimates.

\newpage
\section{Final Remarks and Open Problems}\label{sec:kp-fuzzy:openproblems}

In this article, we were able to provide tight bounds for all values of $\delta$ for both, the additive estimate accuracy for the online simple knapsack with irrevocable decisions, and for the variant with the option to remove items.

In comparison to the setting of the online simple knapsack with estimates and a multiplicative accuracy factor as already covered~\cite{gehnen2024}, it turned out that additive errors are relatively harder to deal with for an algorithm, in particular for the case without removability.
This mainly follows from the fact that, in the additive setting, an adversary can announce items and decide later, when presenting them, if they should be presented as $0$ or with some actual size. As already the pathological counterexample from the beginning has trouble dealing with such items, it is not surprising that all our lower bound constructions consist of multiple tiny items that might turn out to be $0$. In the multiplicative model however such constructions are not possible.

The additive and multiplicative model with removability are much more closely connected to each other, which aligns with our observation about the crucial role of the potentially $0$-items: As in the removability setting all small items can be packed without concerns, as they can be removed anyway, those items turn out to support an online algorithm in its goal of achieving a low competitive ratio. It therefore is not surprising, that an adversary will also not make use of this most crucial difference between the additive and multiplicative setting.

As the variant of online knapsack with estimates can be motivated easily by practical applications,  other variants can be motivated by real-world scenarios as well. Therefore, given some practical problems, it would not be surprising if they could be modeled best with a combination of the estimate model with another knapsack variant:

For example, in the advice setting an algorithm is allowed to ask a limited amount of questions that will be answered truthfully. When you are given the list of estimates, you might again realize that there are just details missing which hinder you from significantly improving your solution quality. Therefore, you might decide to measure a few objects more precisely, as you realize that they can turn out to be crucial.

Connections are also conceivable when additionally allowing to delay some decisions for reservation costs, to exceed the capacity of the knapsack, or to use randomized algorithms. Furthermore, so far we assume that the estimates are correct, e.g., all items are actually close to their estimate. It would be interesting to analyze the behavior in the setting of online algorithms with predictions when the estimates can turn out to be wrong at least partly.

Another approach could be to make an own accuracy $\delta_i$ available for each item $x_i$, instead of just measuring the maximum distortion as we investigated so far.

As for all simple knapsack problems, it would be interesting to extend the setting with estimates to the general knapsack problem. Here several variants are thinkable, for example, which part of the input is predicted, e.g., the size, the weight, or the density of the items.

\newpage

\bibliography{kp-predictions.bib}
\appendix
\end{document}